**Role of anion in the pairing interaction of iron-based superconductivity**

J.-X. Yin[1]*, Y. Y. Zhao[2]*, Zheng Wu[3], X. X. Wu[1], A. Kreisel[4], B. M. Andersen[5], Gennevieve Macam[6], Sen Zhou[7], Rui Wu[1], Limin Liu[1], Hanbin Deng[1], Changjiang Zhu[1], Yuan Li[1], Yingkai Sun[1], Zhi-Quan Huang[6], Feng-Chuan Chuang[6], Hsin Lin[8], C.-S. Ting[3], J.-P. Hu[1], Z. Q. Wang[10], P. C. Dai[11], H. Ding[1], S. H. Pan[1,3, 12,13,14]‡

**Affiliations:**

[1]Institute of Physics, Chinese Academy of Sciences, Beijing 100190, China.

[2]School of Physics and Optoelectronic Engineering, Nanjing University of Information Science and Technology, Nanjing 210044, China.

[3]TCSUH and Department of Physics, University of Houston, Houston, Texas 77204, USA.

[4]Institut für Theoretische Physik, Universität Leipzig, D-04103 Leipzig, Germany.

[5]Niels Bohr Institute, University of Copenhagen, Jagtvej 128, DK-2200 Copenhagen, Denmark.

[6]Department of Physics National Sun Yat-sen University Kaohsiung 80424 Taiwan.

[7]Institute of Theoretical Physics, Chinese Academy of Sciences, Beijing, 100190, China.

[8]Institute of Physics, Academia Sinica Taipei 11529 Taiwan.

[9]Center for Quantum Transport and Thermal Energy Science, Jiangsu Key Lab on Opto-Electronic Technology, School of Physics and Technology, Nanjing Normal University, Nanjing 210097, China.

[10]Department of Physics, Boston College, Chestnut Hill, Massachusetts 02467, USA.

[11]Department of Physics and Astronomy, Rice University, Houston, Texas 77005, USA.

[12]School of Physics, University of Chinese Academy of Sciences, Beijing 100190, China.

[13]CAS Center for Excellence in Topological Quantum Computation, University of Chinese Academy of Sciences, Beijing 100190, China.

[14]Songshang Lake Material Laboratory, Dongguan, Guangdong 523808, China.

*These authors contributed equally to this work.

‡Correspondence to: span@iphy.ac.cn

**High-temperature iron-based superconductivity develops in a structure with unusual lattice-orbital geometry, based on a planar layer of Fe atoms with *3d* orbitals and tetrahedrally coordinated by anions. Here we elucidate the electronic role of anions in the iron-based superconductors utilizing state-of-the-art scanning tunneling microscopy. By measuring the local electronic structure, we find that As anion in $Ba_{0.4}K_{0.6}Fe_2As_2$ has a striking impact on the electron pairing. The superconducting electronic feature can be switched off/on by removing/restoring As atoms on Fe layer at the atomic scale. Our analysis shows that this remarkable atomic switch effect is related to the geometrical cooperation between anion mediated hopping and unconventional pairing interaction. Our results uncover that the local Fe-anion coupling is fundamental for the pairing interaction of iron-based superconductivity, and promise the potential of bottom-up engineering of electron pairing.**

The research on the interplay between geometry and correlation is at the frontier of physics, and often leads to exotic quantum matter (*1-13*). For instance, twisted bilayer graphene at a magic angle introduces emergent flat-band states (*6-8*); interacting spins on the honeycomb lattice can fractionalize into Majorana fermions forming topological spin-liquid (*9,10*); spin-orbit coupled kagome lattice electrons with ferromagnetism can form Chern quantum phases (*11-13*). Other notable examples include high-temperature copper-based superconductors (*2,3*) and iron-based superconductors (*4,5*). While the magnetism of iron would seem to conflict with superconductivity, certain iron-based materials can surprisingly be superconducting at relatively high temperatures. Not only do these superconductors have Fe square lattices, but also their Fe atoms are tetrahedrally coordinated by As (or P, S, Se, Te) anions (*5*). The geometrical anion has long been speculated to mediate the coupling between neighboring Fe ions, tuning the charge and spin degrees of freedom of the system (*4, 14-17*). Structurally, there is evidence that superconductivity has a close relationship with the As-Fe-As bond angle (*18*). However, the electronic role of anions for superconductivity remains elusive at the experimental level.

We use an innovative bottom-up method to investigate this problem by performing atomic layer-resolved scanning tunneling microscopy on different atomic Fe-anion structures. We measure the superconducting electronic variations on these structures, showing the effect of local Fe-anion coupling on the electron pairing. We choose the $Ba_{0.4}K_{0.6}Fe_2As_2$ system (*19*), as it features a high $T_C$ of 38K and can yield multiple terminating layers with our cryogenic cleaving technique (*20*). As surface has a dimerization reconstruction, while Ba(K) and Fe surface has a $\sqrt{2}\times\sqrt{2}R$ 45° reconstruction. Decisive experimental evidence for surface identification was found by imaging the symmetry-dictated surface boundary and the layer-selective chemical dopants (*20*). We thoroughly explore the cleaving surfaces for more than 100 successfully cleaved crystals, and for each cleaved crystal, we scan a large area over 5×5 µm² to search for atomic lattices. Statistically, among their cleaving surfaces, we observe that 75% are disordered surfaces (*21*), 13% are Ba(K) surfaces, 11% are As surfaces, and 1% are Fe surfaces (six samples yeild Fe surfaces). The atomically resolved As surface and Fe surfaces are displayed in Figs. 1**a** and **b**, respectively. We find that the As surface often consists of As dimer vacancies (Fig. 1**a**), and Fe surface often consists of As dimer adatoms (Fig. 1**b**). As vacancies and adatoms can be created by atom manipulation as well. In Fig. 1**c**, we extract an As-dimer from As surface and push it to a clean Fe-surface to demonstrate the formation of a vacancy and an adatom pattern. The As dimer vacancies and adatoms locally break and restore the local tetrahedral anion coordination for Fe ion, respectively, which can affect the local electronic structure and the electron pairing.

Tunneling into As surface, where the Fe-As structure is complete, we observe a fully-opened superconducting gap spectrum with sharp coherence peaks (Fig. 2**a**). The spatial variation of the gap size is less than 7% on clean As surfaces (*22*), and the gap amplitude is consistent with that of Fe $3d_{xz/yz}$ orbital observed by photoemission experiments for the same batch of crystals (*22,23*). The energy range of the flat bottom is shorter than the coherent peak to peak distance, which can be due to the contribution from other orbitals with a smaller energy gap. Studying As dimer vacancy in Fig. 2**b**, we find that spectroscopic imaging displays substantial local variation around the vacancy. The dI/dV mapping at the superconducting gap energy (10meV) in Fig. 2**c** exhibits a pronounced local suppression. The detailed tunneling spectra in Fig. 2**d** reveal that along with the suppression of the superconducting coherence peaks, a pair of sharper in-gap states arise at the vacancy site. The energy of the in-gap state peak is smaller than any superconducting gap detected in photoemission experiments (*22,23*). The in-gap states are bounded to the As vacancy, resembling a local pair-breaking effect associated with this atomic defect (*24*).

Tunneling into Fe surface, where the Fe-As structure is incomplete, we observe a shallow energy gap without coherence peaks (Fig. 2**e**). The shallow incoherent gap supports that superconductivity is strongly suppressed without the As coverage, in agreement with the pair-breaking scenario associated with As vacancy. In addition, the spectrum on Fe surface shows a pair of in-gap shoulders. The in-gap shoulders can be due to the enhanced contribution from other orbitals with a reduced energy gap. At As dimer adatom on Fe surface, the tetrahedral anion coordination can be locally recovered for central Fe ion. Measuring around As dimer adatom in Fig. 2**f**, we find that the spectroscopic imaging at the superconducting gap energy (10meV) exhibits a strong local enhancement of its intensity, as seen from Fig. 2g. Remarkably, As adatom recovers coherence peaks locally (Fig. 2**h**) at the energy the same as that observed on the clean As surface associated with electron pairing of $3d_{xz/yz}$ orbital, while the in-gap shoulder features overall grow deeper.

These dramatic spectral variations are unlikely to arise from a simple tunneling matrix element effect, as the superconducting coherence peak positions can vary slightly near the local Fe-As structures, and the spectrum on Fe surface is gapless. We also perform experiments to confirm that the Fe-As local structures similarly modify the superconducting feature in $Ba_{0.3}K_{0.7}Fe_2As_2$, but at reduced energy scales (6meV) as a result of reduced $T_C$ = 22K; and that the Fe-As local structures do not generate any gap feature in $BaFe_2As_2$ where there is no superconductivity ($T_C$ = 0K). Therefore, the interplays between the local Fe-As structure and the superconducting spectra likely imply that the anions act as an atomic switch of the underlying superconducting order parameter, which is rarely observed in previous tunneling experiments (*25-29*), and we discuss its possible mechanisms below.

Microscopically, the coupling to anions can modify the local electronic structure and affect the electron pairing interaction. The first-principles calculation of the charge distribution in Fe-As structure in Fig. 3**a** shows that the coupling between next-nearest neighbor Fe ions is mediated through As anions (*14,15*), which can crucially modify the electronic structure. Moreover, the widely discussed $s_\pm$ pairing symmetry (*30-32*) in this material can originate from next-nearest neighbor pairing interaction. Therefore, the geometrical match (Fig. 3**b**) between As anion mediated coupling and pairing interaction is essential in the observed pairing phenomena.

To provide a heuristic understanding based on the above discussions, we set up an effective model considering the lattice-orbital geometry of Fe-anion structure (see Supplementary Material) to study the effect of anions on electron pairing. We perform real-space unrestricted self-consistent calculations of the superconducting order parameter and the local density of states (LDOS). We find that by including the local effect of missing As anions on the leading hopping parameters for the $3d_{xz/yz}$ orbital in reference to first-principles calculations, we are able to mimic their essential effects on electron pairing observed in experiments. For As dimer vacancy on As surface, our calculations show that the superconducting coherence peaks are suppressed with the emergence of a pair of in-gap states (Fig. 3**c**). For As dimer adatom on Fe surface, our calculations show that the superconducting coherence peaks disappear on Fe surface and recover locally around As dimer adatom (Fig. 3**d**). Our calculations further uncover the real-space tuning of the electron pairing strength in both cases (Figs. 3**e** and **f**), which support our experimental interpretation that the anions operate as an atomic switch of the superconducting order parameter.

In summary, our experiment uncovers a striking correspondence between local Fe-As structure and superconducting electronic feature in $Ba_{0.4}K_{0.6}Fe_2As_2$, which demonstrates that the Fe-anion coupling is fundamental for iron-based superconductivity, and the electronic impact of anions on electron pairing is local in nature. The novelty of this work is the atomic-scale switch effect of the electron pairing, which is not implied by, or can be derived from, known structural, transport or photoemission effects. Our findings collectively show the rich and unconventional physics of Fe-based superconductivity, which encompasses entangled structure, charge and orbital degrees of freedom, as well as the reformation of electronic states involving many-body electron pairs. A quantitative understanding of this physics would require a comprehensive quantum many-body theory that describes electrons in the Fe-anion lattice in the presence of multi-orbital characteristics and electron-electron interaction. Crucially, our experiment pushes the local effects of electron pairing to frontier discussions (*33*), and imply that the cooperation between lattice-orbital geometry and pairing interaction is fundamental for unconventional superconductivity. Taken together with the atomic manipulability of anions on the Fe surface demonstrated here, our findings also promise the potential to engineer electron pairing with bottom-up construction in the future.

**Acknowledgments:** The authors thank T. Xiang, D.-H. Lee, T-K. Lee, Yi Gao, and J. Kang for stimulating discussions. This work was supported by the Chinese Academy of Sciences, NSFC (Grant No. 11227903, No.11888101), BMSTC (Grant No. Z181100004218007), the Ministry of Science and Technology of China (Grants No. No. 2015CB921304, and No. 2017YFA0302903), the Strategic Priority Research ProgramB (Grants No. XDB04040300 and No. XDB07000000). Work at University of Houston was supported by the State of Texas through TcSUH. Y.Y.Z. is supported by the National Natural Science Foundation of China (Grant No. 11804163), the Natural Science Foundation of the JiangSu Higher Education Institutions of China (Grant No. 18KJB140006). C.-S.T. acknowledges support of the Robert A. Welch Foundation (Grant No. E-1146). Z.W. is supported by the U.S. Department of Energy, Basic Energy Sciences (Grant No. DE-FG02-99ER45747). B. M. A. acknowledges support from the Independent Research Fund Denmark grant number 8021-00047B.

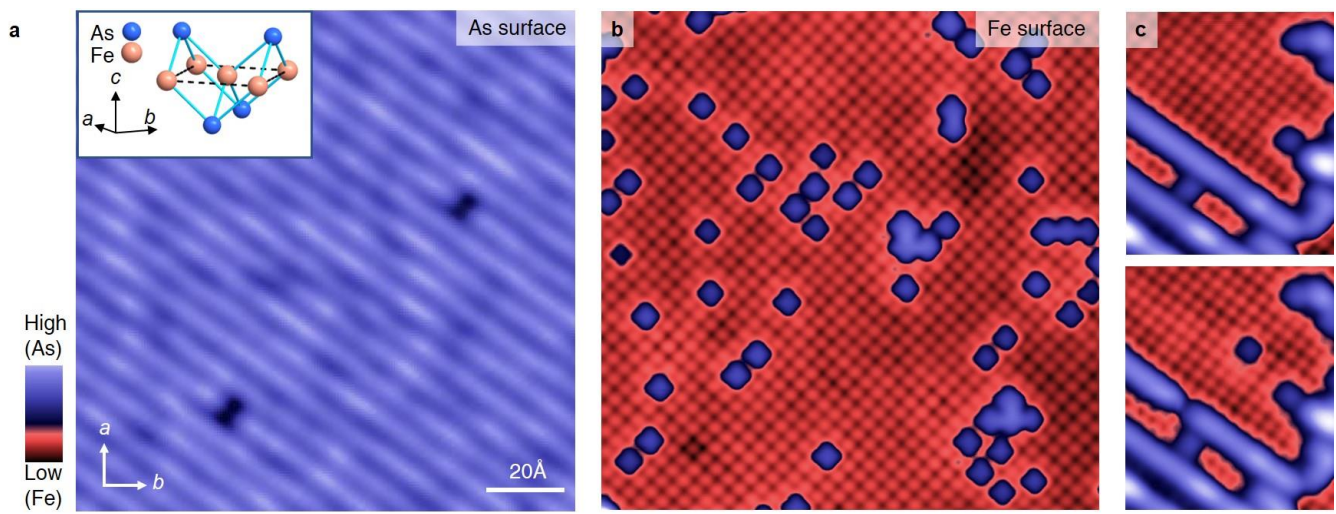

**Fig. 1 Atomic-scale Fe-As structures. a,** A topographic image showing As surface with As dimer vacancies. The inset shows the crystal structure of the Fe-As unit cell. **b,** A topographic image displaying Fe surface with As adatoms. **c,** Extracting an As-dimer from As surface and moving it to a clean Fe-surface. The upper and lower panels show the topographic image before and after the atom manipulation, respectively.

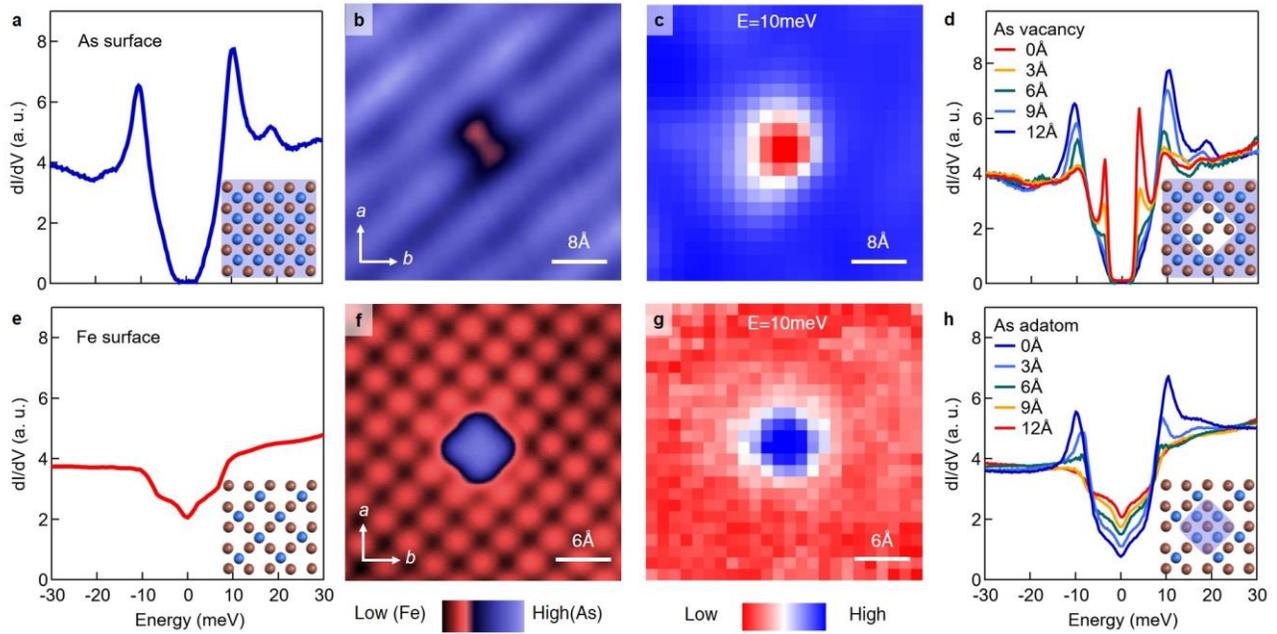

**Fig. 2. Impact of Fe-As coupling on the superconducting spectrum. a,** Tunneling spectrum taken at As surface. The inset shows the corresponding Fe-As structure. **b,** Topographic image of As dimer vacancy. **c,** dI/dV map taken at 10meV for the As vacancy. **d,** Tunneling spectrum taken from the center of As dimer vacancy to a faraway position. The inset shows the corresponding Fe-As structure. **e,** Tunneling spectrum taken at the Fe surface. The inset shows the corresponding Fe-As structure. **f,** Topographic image of As dimer adatom on the Fe surface. **g,** dI/dV map taken at 10meV for As dimer adatom. **h,** Tunneling spectrum taken from the center of As dimer adatom to a faraway position. The inset shows the corresponding Fe-As structure.

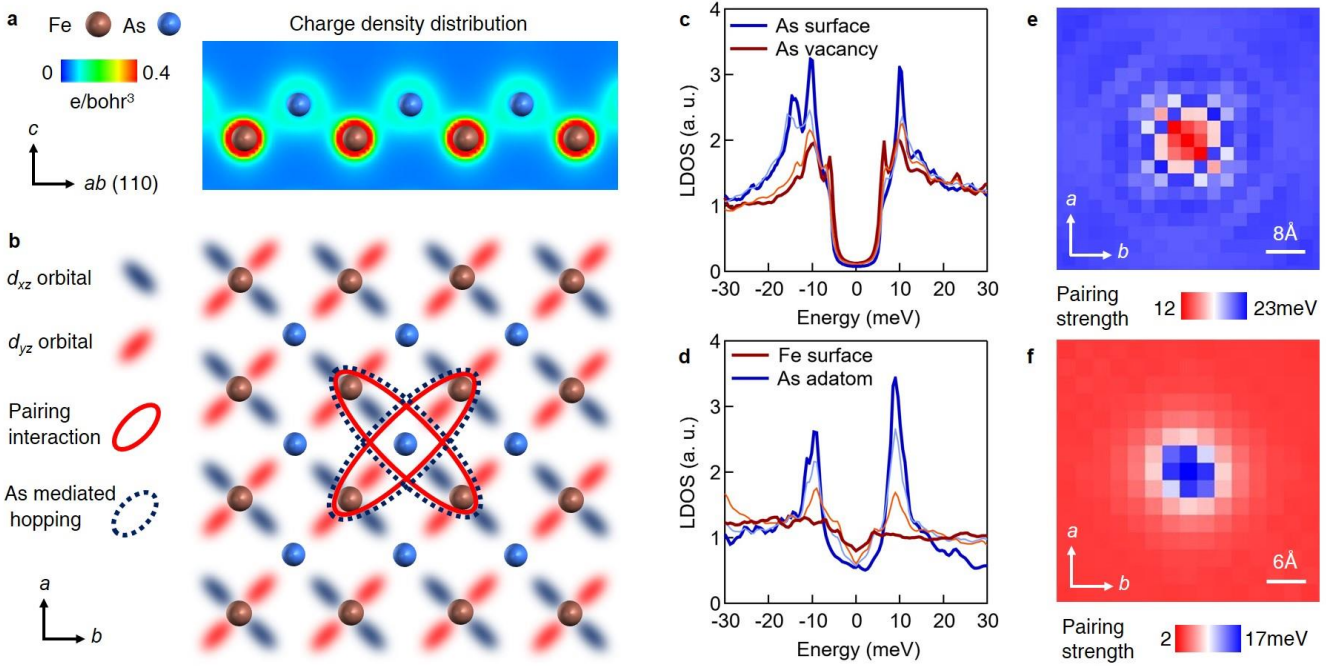

**Fig. 3. Modeling of local Fe-As coupling on superconductivity. a,** First-principles calculated charge-density distribution of Fe-As structure in $BaFe_2As_2$ from the side view, showing that the coupling between next-nearest neighboring Fe ions is through As anions. **b,** Lattice-orbital geometry of Fe-As structure from the top view. We consider next-nearest neighbor pairing interaction leading to s± pairing symmetry, and each As anion essentially mediates the hopping between nest-nearest neighbor Fe ions. **c,** Calculated LDOS for As dimer vacancy on the As surface. The thinner lines are representative LDOS near As vacancy. **d,** Calculated LDOS for As dimer adatom on the Fe surface. The thinner lines are representative LDOS near As adatom. **e,** Calculated pairing strength (order parameter) map for As dimer vacancy on an As surface. **f,** Calculated order parameter map for As dimer adatom on a Fe surface. All the calculations are performed on a 32×32 square lattice with periodic boundary conditions.

## Materials and Methods

The single crystalline samples of $Ba_{0.4}K_{0.6}Fe_2As_2$ were grown using the self-flux method. The scanning tunneling microscopy experiments were carried out on a home-built ultrahigh vacuum low-temperature scanning tunneling microscope. Samples were cleaved in-situ below 30 K and immediately transferred to the microscope head, which was already at the base temperature of 4.2 K. The scan tips were mainly prepared from polycrystalline tungsten wires by electrochemical etching and subsequent field-emission cleaning. Identical spectra on the As adatom and As vacancy were observed by using a Pt/Ir tip. Topographic images were acquired in the constant-current mode with the bias voltage applied to the sample, with a typical voltage of 100mV and tunneling current of 0.1nA. Differential conductance spectra were recorded with the standard lock-in technique with energy modulation of 0.2meV at a temperature from 1.5K to 25K.

## Extended topographic data on the Fe surface

Figure. S1a shows the topography of Fe surface with a large field of view. Both individual As dimer adatoms and As dimer rows are observed on the surface. Based on the crystalline symmetry of As-Fe-As trilayer structure and surface atom reconstructions, we illustrate the details of the atomic structure of As vacancy and As adatom associated with the atom manipulation shown in the main text in Fig. S1b.

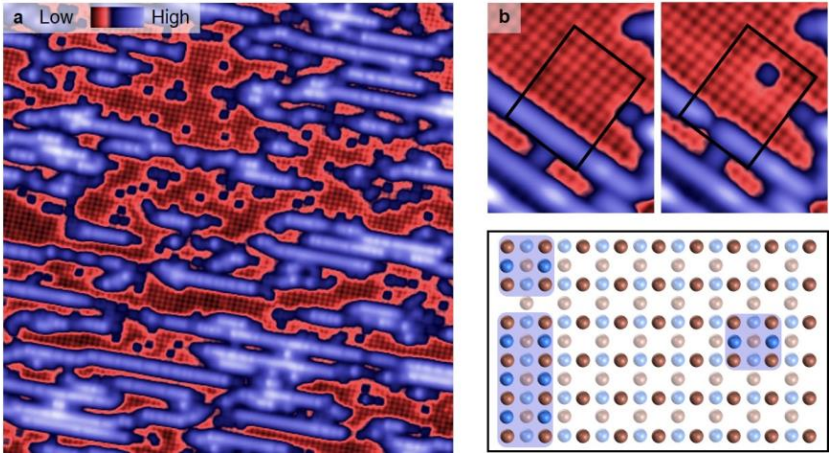

**Fig. S1. Extended topographic data on the Fe surface. a,** A topography of exposed Fe surface showing both As dimer adatoms and As dimer rows (300Å × 300Å). **b,** Atomic manipulation (reproduced from Fig. 1**c**, top panels) and the corresponding schematic (lower panel). The dark and light red balls represent Fe atoms with $\sqrt{2} \times \sqrt{2}\ R45°$ surface reconstruction with half of Fe atoms invisible (*20*). The dark and light blue balls represent the As atoms above and below the Fe lattice, respectively.

## Realistic theoretical modeling

### Basic model with As-Fe-As lattice-orbital geometry

Iron-based superconductors have been theoretically investigated for many years (*5, 30-32, 34-37*), and several tentative effective models (*38-48*) have been proposed to study the electronic correlations of the system. In order to explain the experimental results above, we choose a two-orbital tight-binding model that provides the key symmetry and phase diagram of hole-doped (Ba,K)Fe$_2$As$_2$ compounds (*48-53*). In the model, $t_1$ and $t_5$ represent the intra- and inter- nearest-neighbor (NN) hopping integrals, respectively. $t_2$ and $t_3$ represent the intra- next-nearest-neighbor (NNN) hopping integrals, while $t_4$ represents the inter- NNN hopping integral. We consider the $s_\pm$- pairing symmetry and choose the NNN intraorbital pairing with strength $V_{i\mu j\mu} = V_{ij} = V_{NNN}$ as a constant. Local magnetization $m_i$ and $s_\pm$- pairing projection of the superconductivity order parameter $\Delta_i$ at each site $i$ can be calculated, respectively. Throughout our calculations, the real-space numerical calculations are performed on a $32 \times 32$ square lattice with periodic boundary conditions. The input parameters are the same as those in Ref. 48.

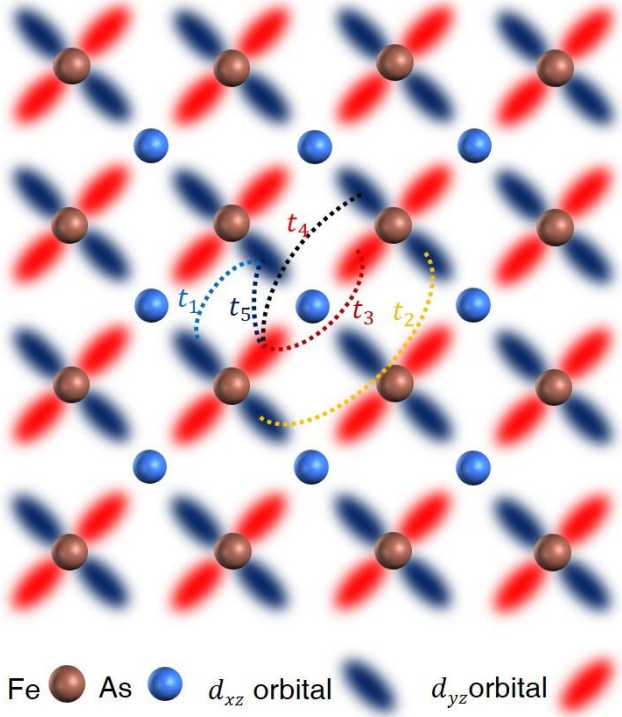

**Fig. S2**. **Schematics showing the electron hopping.** The leading hopping terms $t_5$ and $t_3$ are affected by As anions in our consideration.

**Realistic minimum modeling of missing As anions**

To minimize the modeling of our experiments, we consider that the leading NNN hopping parameter $t_3$ and leading NN hopping parameter $t_5$ can be altered by the missing As anions. As a reference, we estimate the hopping amplitudes of $t_3$ and $t_5$ following the Slater-Koster formalism (*54,41*). The first-principles calculation suggests that $|t_3|$ is substantially reduced from 0.34eV to 0.04eV, while $|t_5|$ is slightly enhanced from 0.12eV to 0.15eV when removing the orbital hybridization from As $p$ orbitals. This calculation is consistent with the previous theory (*14,15*) and guides us to change $t_3$ and $t_5$ in our model. To further constrain these two parameters, we calculate the band structure of BaFe$_2$As (without one As anion layer) with our model and first-principles in Fig. S3. To reproduce the key trend and symmetry of band dispersion in first-principles calculations for our model, we find that $t_3$ needs to be substantially reduced from 1.35 to 0.04±0.06 and $t_5$ needs to be slightly enhanced from -1 to -1.3±0.1.

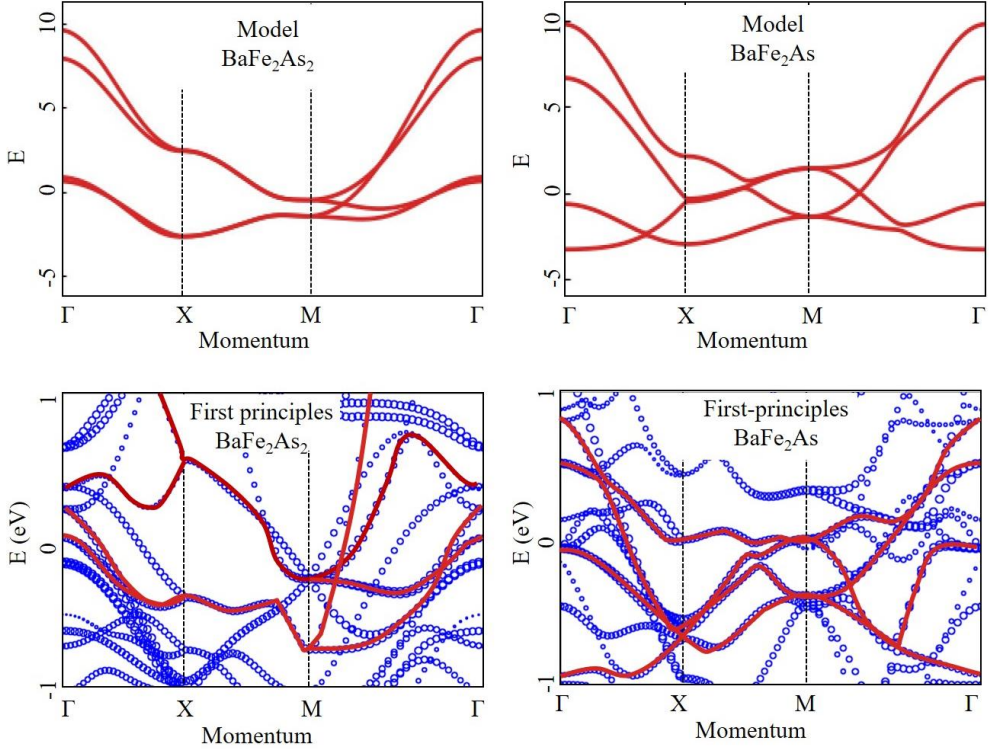

**Fig. S3 Calculated band structure comparison.** The two images show the model calculated band structures for $BaFe_2As_2$ and $BaFe_2As$ (without one As anion layer), respectively. The lower images show the first-principles calculated band structure for $BaFe_2As_2$ and $BaFe_2As$, respectively. For the lower images, the blue circles mark the spectral weight of Fe $3d$ orbitals, and the red curves illustrate the key band structure corresponding to the model scenarios (isomorphic bands with the model).

**Energy units and phase diagram**

From Fig. S3 it can be inferred that the bandwidth of the model is 13, and the related band in the first-principles calculation has a bandwidth around 2eV. Based on the comparison between first-principles and measured band structure from angle-resolved photoemission data in $(Ba,K)Fe_2As_2$ system (*55*), the band renormalization factor is 2. Therefore, the model bandwidth corresponds to 1eV in real materials, and we can assign real energy scale units based on this ratio (1 in the model corresponds to 0.077eV). We also compare the model calculated phase diagram (*48*) with the transport data in $(Ba,K)Fe_2As_2$ (*56*) in Fig. S4, which shows reasonable consistency. The phase diagram consistency supports us in applying the model to $Ba_{0.4}K_{0.6}Fe_2As_2$ in our scanning tunneling microscopy experiment, which is illustrated in Fig. S5 for the real space calculations.

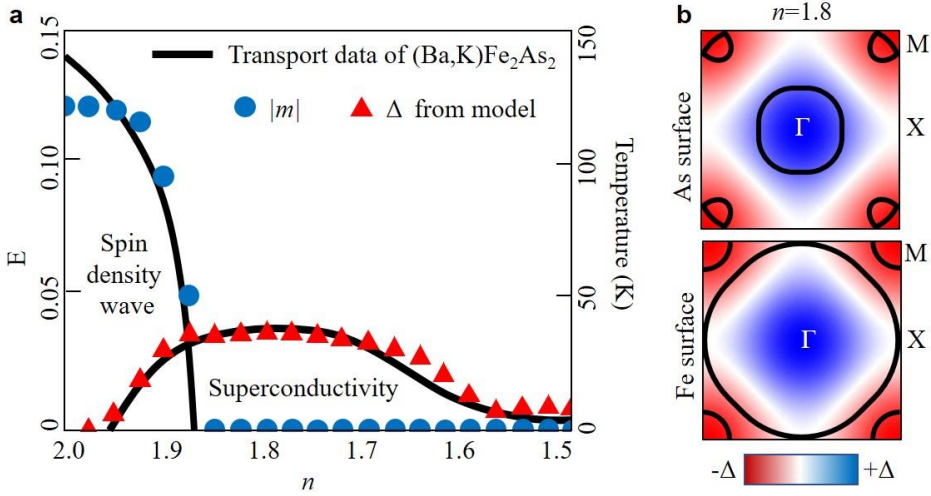

**Fig. S4. Model calculated phase diagram and Fermi surfaces. a,** Model calculated phase diagram in comparison with transport data. **b,** Model calculated Fermi surfaces for As surface and Fe surface in comparison with the gap anisotropy of $s_\pm$ wave. The Fermi surface for Fe surface overlaps with the nodal line of $s_\pm$ wave, leading to gapless superconductivity in the model.

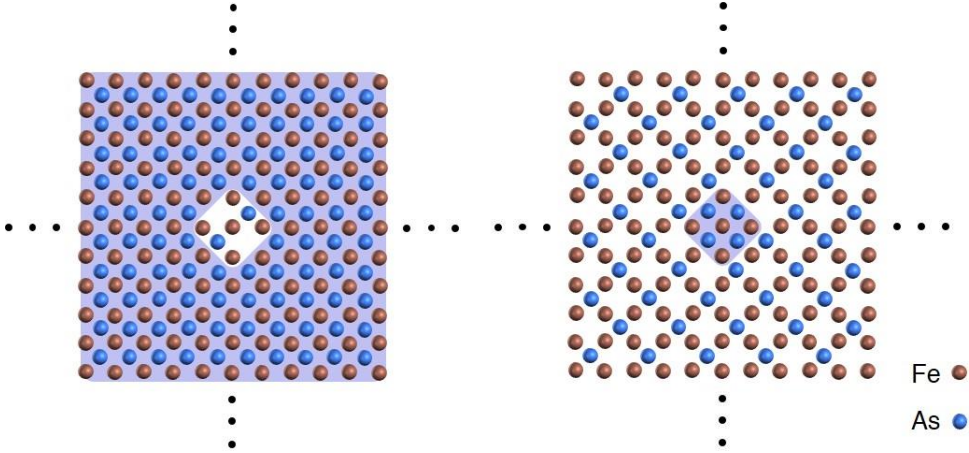

**Fig. S5. Schematic images for model calculation set up for As dimer vacancy on As surface (left) and As dimer adatom on Fe surface (right).**

## First-principles calculation

First-principles density functional theory calculations (*57,58*) were performed using the Vienna Ab initio Simulation Package (*59,60*). The Perdew-Burke-Ernzerhof generalized gradient approximation functional (*61*) was applied to describe electron exchange-correlation interaction with the projected augmented wave potentials (*62*). The energy cutoff was set to 300 eV and the criteria for convergence of non-spin-polarized self-consistent calculations were set to $10^{-5}$ eV. Atomic structures were optimized until the residual forces have converged to less than $10^{-3}$ eV/Å. The Brillouin zone was sampled using a 24×24×8 Monkhorst Pack grid (*63*). To study the effect of As anions on the hoppings between $d$ orbitals, we perform calculations for the bulk BaFe$_2$As$_2$ and the energy cutoff was set to be 500 eV and the k mesh is 7×7×7 for the primitive cell. We used the maximally localized Wannier functions to construct tight-binding models by fitting the first-principles band structure. The hoppings between Fe orbitals can be calculated as $t_d^{eff} = t_{dd} + t_{dpd}$, where the first term is the direct hopping and the second term is the hopping through As atoms (*54*, *41*).